\documentclass[reprint,aps,prd,nofootinbib,floatfix]{revtex4-1}
\usepackage{ulem}
\usepackage{amsmath}
\usepackage{graphicx}
\usepackage{subfigure}
\usepackage{multirow}
\usepackage{color}
\usepackage{enumitem}
\begin{document}
\title{$\gamma$-Ray Puzzles in Cygnus X: Implications for High-Energy Neutrinos}

\author{Tova M. Yoast-Hull$^{1,2}$, John S. Gallagher III$^3$, Francis Halzen$^{1,2}$, Ali Kheirandish$^{1,2}$, and Ellen G. Zweibel$^{1,3}$}
\affiliation{$^1$Department of Physics, University of Wisconsin-Madison, Madison, WI, USA, 53706}
\affiliation{$^2$Wisconsin IceCube Particle Astrophysics Center, University of Wisconsin-Madison, Madison, WI, USA, 53703}
\affiliation{$^3$Department of Astronomy, University of Wisconsin-Madison, Madison, WI, USA, 53706}

\begin{abstract}
The Cygnus X region contains giant molecular cloud complexes and populous associates of massive young stars. The discovery of spatially extended, hard $\gamma$-ray emission in Cygnus X by both Milagro and \textit{Fermi} indicates that Cygnus X is also a potential source of high-energy Galactic neutrinos. Here, we adapt our single-zone model for cosmic ray interactions in the central molecular zones of starburst galaxies for use in Cygnus X. We calculate the potential neutrino flux corresponding to the hard $\gamma$-ray emission from the ``Cygnus Cocoon" and to the soft, diffuse interstellar $\gamma$-ray emission. We check our results by comparing the corresponding $\gamma$-ray emission against the \textit{Fermi} interstellar emission model and Milagro, ARGO-YBJ, and HAWC observations.  In comparing our results against a recent IceCube analysis and the current sensitivity limits, we find that neutrino emission from the Cocoon has a large enough flux that it could plausibly be detected, provided hadronic interactions are occurring at sufficiently high energies.  High-energy neutrinos from Cygnus X would provide direct evidence for the presence of as yet unidentified PeV energy accelerators in the Galactic disk.
\end{abstract}

\maketitle

\section{Introduction}

The detection of high-energy astrophysical neutrinos by IceCube has opened a new window into cosmic ray astrophysics and a new path for studies of potential cosmic particle accelerators operating at PeV energies \citep[e.g.,][]{Aartsen14, Aartsen16, Aartsen17a}.  Unlike their high-energy cosmic ray counterparts, neutrinos can pass through galactic and intergalactic magnetic fields without changing their direction, and as such, neutrinos can be traced back to their original sources. This combined with the relatively low rates of interaction with intervening materials allows sources with high cosmic-ray hadronic number densities to be identified.

The detection of TeV energy $\gamma$-ray sources in the Galactic plane by Milagro and ARGO-YBJ provides potential clues for finding high-energy neutrino sources and cosmic accelerators \citep{Halzen08}. Energetic sources in the Galactic plane are likely to be in the Milky Way and, thus, sufficiently nearby to be studied in detail. In a recent review of several Galactic TeV $\gamma$-ray sources, \citep{Halzen17}, the authors note that MGRO~J2031+41, which is spatially coincident with the well-studied Cygnus~X complex, is promising as a nearby source of high-energy neutrinos that can be detected by IceCube \citep{Ackermann11,Aharonian02,Anchordoqui09,Grenier13,Tchernin13,Gonzalez14,Aartsen15,Niers15,Halzen17}.

The Cygnus~X region is a nearby ($D = 1.3$~kpc) example of a giant star-forming complex containing massive molecular gas clouds, rich populations of young stars, and luminous H\textsc{II} regions \citep{Baars81, Leung92, Kiminki15}.  Cygnus~X is a bright source of $\gamma$-rays containing both soft and hard spatially extended components and multiple point sources, including supernova remnants and pulsars \citep{Ackermann11}.  The presence of a hard $\gamma$-ray spectrum, in combination with dense molecular clouds \citep{Gottschalk12, Schneider16} and a large number of young OB stars \citep{Wright15}, suggests that the Cygnus~X region could be a source of recently accelerated cosmic rays and high-energy astrophysical neutrinos \citep{Aharonian02, Ackermann11, Grenier13, Tchernin13, Gonzalez14, Aartsen15, Niers15}.

The Cygnus~X region was tentatively detected in $\gamma$-rays by EGRET \citep{Chen96}, and this observation was later used to confirm the Cygnus~X region as a source differing from the Cyg X-3 binary by \citep{Mori97}.  Hard $\gamma$-ray emission from the region was confirmed with HEGRA by \citep{Aharonian02}.  Properties of the $\gamma$-rays from the Cygnus~X region have since been extensively explored at both GeV and TeV energies with Milagro \citep{Abdo07}, MAGIC \citep{Albert08}, \textit{Fermi} \citep{Ackermann11, Ackermann12}, ARGO-YBJ \citep{Bartoli14}, and VERITAS \citep{Aliu14, Popkow15}.  

The differing fields of view and energy discrimination between the various $\gamma$-ray detectors and the combinations of point sources and extended emission all make interpretation of the $\gamma$-ray data complex.  Additionally, near-infrared observations of the stellar populations of Cygnus~X reveal a significant range of ages among the young stars, which suggests that there may be unidentified accelerators and sources, such as pulsar wind nebulae (PWNe) or supernova remnants (SNRs), in the region \citep{Comeron16}.  Aside from the many point sources present in Cygnus~X, another important feature of $\gamma$-ray observations of the region is the hard, extended emission, referred to as the Cygnus Cocoon by the \textit{Fermi} collaboration.  

The Cocoon, defined as `an extended excess of hard emission above the modeled background,' was first detected by \citep{Ackermann11} by subtracting out the isotropic $\gamma$-ray background, point sources in the region, and the modeled interstellar $\gamma$-ray radiation from the total $\gamma$-ray emission from the Cygnus region.  Further, extended emission has been detected at TeV energies (by Milagro, ARGO-YBJ, VERITAS, HAWC) that is spatially coincident with small portions of the Cocoon.  It has yet to be established whether the Cocoon is a single entity, potentially coming from a region covering $\sim 10$ deg$^{2}$ on the sky, or some combination of unresolved point sources and smaller regions of extended emission.

In this paper, we develop models for the possible neutrino fluxes from Cygnus~X based on the spatially extended $\gamma$-ray observations.  We derive gas column densities for atomic, molecular, and neutral hydrogen gas from recent \textit{Planck} observations of the Galactic plane and calculate the interstellar radiation field in Cygnus~X from IRAS 100 micron observations.  Combining this model for the interstellar medium with local cosmic ray observations \cite{Vos15}, we calculate the soft, extended $\gamma$-ray component in Cygnus~X and compare our findings with the Galactic Interstellar Emission Model (GIEM) adopted by the Large Area Telescope (LAT) Collaboration.  We also calculate the neutrino flux and compare the results with the current IceCube sensitivity limits.  Finally, we compute an upper limit for the possible neutrino flux from the giant molecular cloud in CygX-North and from the collective Cygnus Cocoon, assuming hadronic $\gamma$-ray emission only.

\section{Model Setup}

\subsection{Theoretical Approach}

\subsubsection{Primary and Secondary Cosmic Rays}

For simplicity's sake, we begin by assuming that the cosmic ray spectrum observed at Earth is representative of a cosmic ray spectrum distributed uniformly throughout the galaxy.  A parametrization of this spectrum, fit to observations from Voyager, AMS, and Pamela, is given by \citep{Vos15}
\begin{equation}\label{proton}
N_{p}(T_{p}) = 1.08 \pi \frac{T_{p}^{1.12}}{c \beta^{2}} \left( \frac{T_{p} + 0.67}{1.67} \right)^{-3.93},
\end{equation}
where $T_{p}$ is the kinetic energy of the cosmic-ray proton. Similarly, a parameterization of the cosmic-ray electron spectrum is given by \citep{Potgieter15}
\begin{equation}\label{electron}
N_{e}(T_{e}) = 0.084 \pi \frac{T_{e}^{-1.35}}{c \beta^{2}} \left( \frac{T_{e}^{1.65} + 0.6920}{1.6920} \right)^{-1.1515}.
\end{equation}
Both equations are given in units of particles cm$^{-3}$ GeV$^{-1}$.

Inclusion of secondary cosmic rays produced in proton-proton interactions is critical to accurately modeling Cygnus X as the region is known to contain molecular clouds with high column densities (e.g., $N \sim 10^{22}$ cm$^{-2}$ \citep{Schneider06}).  The main products of proton-proton interactions are charged and neutral pions.  The source function for these pions depends on the ISM density ($n_{\text{ISM}}$) and the proton energy spectrum ($N_{p}$) such that
\begin{equation}\label{pion}
q_{\pi}(E_{\pi}) = \frac{cn_{\text{ISM}}}{K_{\pi}} ~ \xi_{\pi}(E_{p}) \sigma_{\text{inel}}(E_{p}) N_{p}(T_{p}),
\end{equation}
where $\xi_{\pi}(E_{p})$ is the pion multiplicity and $E_{p} = m_{p}c^{2} + T_{p} = m_{p}c^{2} + E_{\pi} / K_{\pi}$, with $K_{\pi} \approx 0.17$ being the fraction of proton kinetic energy transferred to the resulting pion \citep{Kelner06}.

For charged pions, the multiplicity, $\xi_{\pi}$, is taken from the ratio of the inclusive cross sections found in \citep{Dermer86} such that $\xi_{\pi^{\pm}}(E_{p}) = \sigma_{\pi^{\pm}}(E_{p}) / \sigma_{\pi^{0}}(E_{p})$ \citep{Murphy87}.  For neutral pions, the multiplicity is merely $\xi_{\pi^{0}} = 1$.

As pions are relatively short lived particles, it is their decay products (electrons, positrons, neutrinos, and $\gamma$-rays) which we will focus on.  Charged pions decay into charged muons which subsequently decay into secondary electrons and positrons.  The source function for secondary electrons and positrons is given by
\begin{equation}\label{electron1}
q_{e^{\pm}}(\gamma_{e}) = \frac{m_{\mu}}{m_{e}} \int_{1}^{B} d\gamma_{e}' \frac{P(\gamma_{e}')}{2\sqrt{\gamma_{e}'^{2}-1}} \int_{\gamma_{\mu}^{-}}^{\gamma_{\mu}^{+}} d\gamma_{\mu} \frac{q_{\mu^{\pm}}(\gamma_{\mu})}{\sqrt{\gamma_{\mu}^{2}-1}},
\end{equation}
where\footnote{Note that the factor of $m_{\mu} / m_{e}$ in this equation does not appear in the cited texts as the original units for $q(\gamma)$ were $\gamma^{-1}$ cm$^{-3}$ s$^{-1}$ instead of GeV$^{-1}$ cm$^{-3}$ s$^{-1}$.} $\gamma_{\mu}^{\pm} = \gamma_{e}\gamma_{e}' \pm \sqrt{\gamma_{e}^{2}-1} \sqrt{\gamma_{e}'^{2}-1}$, $B = m_{\mu} / 2 m_{e} \sim 104$, and the electron/positron distribution in the muon's rest frame is $P(\gamma_{e}') = 2 \gamma_{e}'^{2} ( 3 - 2 \gamma_{e}' / B) / B^{3}$ \citep{Jones63, Ramaty66, Ginzburg69, Schlick02}.  The similar rest masses of pions and muons allows us to make the substitution $q_{\pi}(\gamma_{\pi}) = q_{\mu}(\gamma_{\mu})$ \citep{Schlick02}.

Because the region in which the secondary cosmic rays are produced is the same as the interaction region, we can take advantage of the approximation that is used in our semianalytic modeling approach in \citep{YoastHull13}, hereafter known as the YEGZ models.  Thus,
\begin{equation}\label{secondary}
N(E) \approx Q(E) \tau(E).
\end{equation}
The cosmic ray lifetime includes advection and diffusion timescales and an additional energy loss lifetime such that $\tau^{-1} = \tau_{\text{loss}}^{-1} + \tau_{\text{adv}}^{-1} + \tau_{\text{diff}}^{-1}$.  These timescales are given by
\begin{align}\label{lifetime}
\tau_{\text{loss}}(E) &= - \frac{E}{dE/dt},\\
\tau_{\text{adv}} &= \frac{d}{v_{\text{adv}}},\\
\tau_{\text{diff}}(E) &= \frac{d^{2}}{3 D_{0}} \left( \frac{B / 3 \mu\text{G}}{E / 1\text{GeV}} \right)^{\beta}.
\end{align}
Energy loss mechanisms for cosmic ray electrons include ionization, bremsstrahlung, inverse Compton, and synchrotron; the energy loss rates can be found in \citep{YoastHull13}.  Due to the high gas densities and photon energy densities found in Cygnus X, the effects of both advection and diffusion are negligible for secondary electrons and positrons.  Thus, the timescale for secondaries is effectively reduced to the energy loss lifetime.

Our advection timescale is assumed to be energy independent, where $d$ is the depth of the region and the wind advection speed is assumed to be $v_{\text{adv}} = 50$ km~s$^{-1}$, which is within a factor of a few of the Alfv\'{e}n speed of the cosmic rays.  In regards to the diffusion timescale, we assume an diffusion coefficient of $10^{28}$ cm~s$^{-1}$ and a spectral index of $\beta = 1/3$, consistent with scattering by a Kolmogorov spectrum of turbulence.  Also, note that we assume a standard galactic magnetic field strength of 5~$\mu$G throughout the region \citep{Ferriere01}.

\subsubsection{Gamma-Rays and Neutrinos}

As noted above, charged pions decay into muons and subsequently into secondary electrons and positrons.  In conserving the lepton number of these weak interactions, two muon neutrinos and an electron neutrino are also produced: $\pi \rightarrow \mu + \nu_{\mu}^{(1)}$ and $\mu \rightarrow e + \nu_{\mu}^{(2)} + \nu_{e}$.

Because three-particle decays are quite complex, we use the analytical approximations, found in \citep{Kelner06}, which are based on the \textsc{SYBILL} code for secondary particles with energies above 100 GeV.  Based on Eqs. (71) and (72) in \citep{Kelner06}, the neutrino emissivity can be represented as
\begin{equation}
q_{\nu}(E_{\nu}) = c n_{\text{ISM}} \int_{0}^{1} F_{\nu} \left(x, \frac{E_{\nu}}{x} \right) \sigma_{\text{inel}} \left(\frac{E_{\nu}}{x} \right) N_{p} \left(\frac{E_{\nu}}{x} \right) \frac{dx}{x},
\end{equation}
where $x = E_{\nu} / E_{\pi}$ and $F_{\nu} = F_{\nu_{\mu}}^{(1)} + F_{\nu_{\mu}}^{(2)} + F_{\nu_{e}}$.  Expressions for each of the neutrino distribution functions $F_{\nu_{\mu}}^{(2)}$ and $F_{\nu_{e}}$ can both be approximated as $F_{e}$.  The equations for $F_{\nu_{e}}$ and $F_{\nu_{\mu}}^{(1)}$ correspond to Eqs. (62)--(65) and Eqs. (66)--(69) in \citep{Kelner06}.

In addition to secondary cosmic rays and neutrinos, $\gamma$-rays also result from the decay of pions from proton-proton interactions.  To conserve momentum, neutral pions decay in to two $\gamma$-rays and the emissivity for this process is given by \citep{Stecker71}
\begin{equation}
q_{\gamma,\pi^{0}}(E_{\gamma}) = 2 \int_{E_{\text{min}}}^{\infty} dE_{\pi} \frac{q_{\pi}(E_{\pi})}{\sqrt{E_{\pi}^{2} - m_{\pi}^{2}c^{4}}},
\end{equation}
where $E_{\text{min}} = E_{\gamma} + m_{\pi}^{2}c^{4} / (4 E_{\gamma})$.  

To calculate the total $\gamma$-ray spectrum, we must also include leptonic production processes and account for the combined spectrum from both primary electrons and secondary electrons and positrons, $N_{e}(E_{e}) = N_{e-}^{\text{prim}}(E_{e}) + N_{e-}^{\text{sec}}(E_{e}) + N_{e+}^{\text{sec}}(E_{e})$  [in units of cm$^{-3}$ GeV$^{-1}$], following from Eqs.~(\ref{proton}), (\ref{electron1}), (\ref{secondary}), and (\ref{lifetime}).  Cosmic-ray electrons and positrons produce $\gamma$-rays via bremsstrahlung in their interactions with the ISM.  The emissivity for $\gamma$-rays from bremsstrahlung is given by
\begin{equation}
q_{\gamma,\text{Br}} (E_{\gamma}) = c n_{\text{ISM}} \int_{E_{\text{min}}}^{\infty} dE_{e} ~ N_{e}(E_{e}) \frac{d\sigma}{dE_{\gamma}},
\end{equation}
where $E_{\text{min}} = \sqrt{E_{\gamma} ( 2 m_{e}c^{2} + E_{\gamma})}$ \citep{Blumenthal70}.  The differential cross section is given by
\begin{equation}
\frac{d\sigma}{dE_{\gamma}} = \frac{3 \alpha \sigma_{T}}{8 \pi E_{\gamma}} \left[ \left[ 1 + \left( 1 - \frac{E_{\gamma}}{E_{e}} \right)^{2} \right] \phi_{1} - \frac{2}{3} \left( 1 - \frac{E_{\gamma}}{E_{e}} \right) \phi_{2} \right],
\end{equation}
where the scattering functions $\phi_{1} = \phi_{2} = Z^{2} \phi_{u}$ and $\phi_{u}$ is given by \citep{Ginzburg69, Blumenthal70, Dermer09}
\begin{equation}
\phi_{u} = 4 \left[ \text{ln} \left( \frac{2E_{e}}{m_{e}c^{2}} \left( \frac{E_{e} - E_{\gamma}}{E_{\gamma}} \right) \right) - \frac{1}{2}  \right].
\end{equation}

Lastly, interactions between cosmic-ray leptons and interstellar radiation, primarily infrared and starlight, result in $\gamma$-rays via inverse Compton.  The inverse Compton $\gamma$-ray emissivity is given by
\begin{equation}
q_{\gamma,\text{IC}}(E_{\gamma}) = \frac{3 c \sigma_{T}}{16 \pi} \int_{0}^{\infty} d\epsilon \frac{v(\epsilon)}{\epsilon} \int_{E_{\text{min}}}^{\infty} dE_{e} \frac{N_{e}(E_{e})}{\gamma_{e}^{2}} F(q, \Gamma).
\end{equation}
The minimum cosmic ray energy is given by \citep{Schlick02}
\begin{equation*}
E_{\text{min}} = \frac{1}{2} E_{\gamma} \left[ 1 + \left( 1 + \frac{m_{e}^{2}c^{4}}{\epsilon E_{\gamma}} \right)^{1/2} \right],
\end{equation*}
where $E_{\gamma}$ is the energy of the resulting $\gamma$-ray, $\epsilon$ is the energy of the incident photon, and $E_{e}$ is the energy of the electron.  The function $F(q, \Gamma)$ is part of the Klein-Nishina cross section and is given by \citep{Blumenthal70}
\begin{equation*}
F(q, \Gamma) = 2q \text{ln}(q) + (1 + q - 2q^{2}) + \frac{\Gamma^{2}q^{2} (1 - q)}{2(1 + \Gamma q)},
\end{equation*}
where
\begin{equation*}
\Gamma = \frac{4 \epsilon \gamma_{e}}{(m_{e}c^{2})} \text{ and } q = \frac{E_{\gamma}}{\Gamma (\gamma_{e} m_{e}c^{2} - E_{\gamma})}.  
\end{equation*}
For the infrared and stellar radiation fields, we assume an isotropic, diluted, modified blackbody spectrum \citep{Casey12}
\begin{equation}
v(\epsilon) = \frac{C_{\text{dil}}}{\pi^{2} \hbar^{3} c^{3}} \frac{\epsilon^{2}}{e^{\epsilon / k T} - 1} \left( \frac{\epsilon}{\epsilon_{0}} \right)^{1.6},
\end{equation}
where $C_{\text{dil}}$ is a spatial dilution factor (given by the normalization $U_{\text{rad}} = \int v(\epsilon) \epsilon d\epsilon$) and $\epsilon_{0}$ corresponds to $\lambda_{0} = 200 ~ \mu$m.

\begin{figure*}
 \subfigure{
  \includegraphics[width=\linewidth]{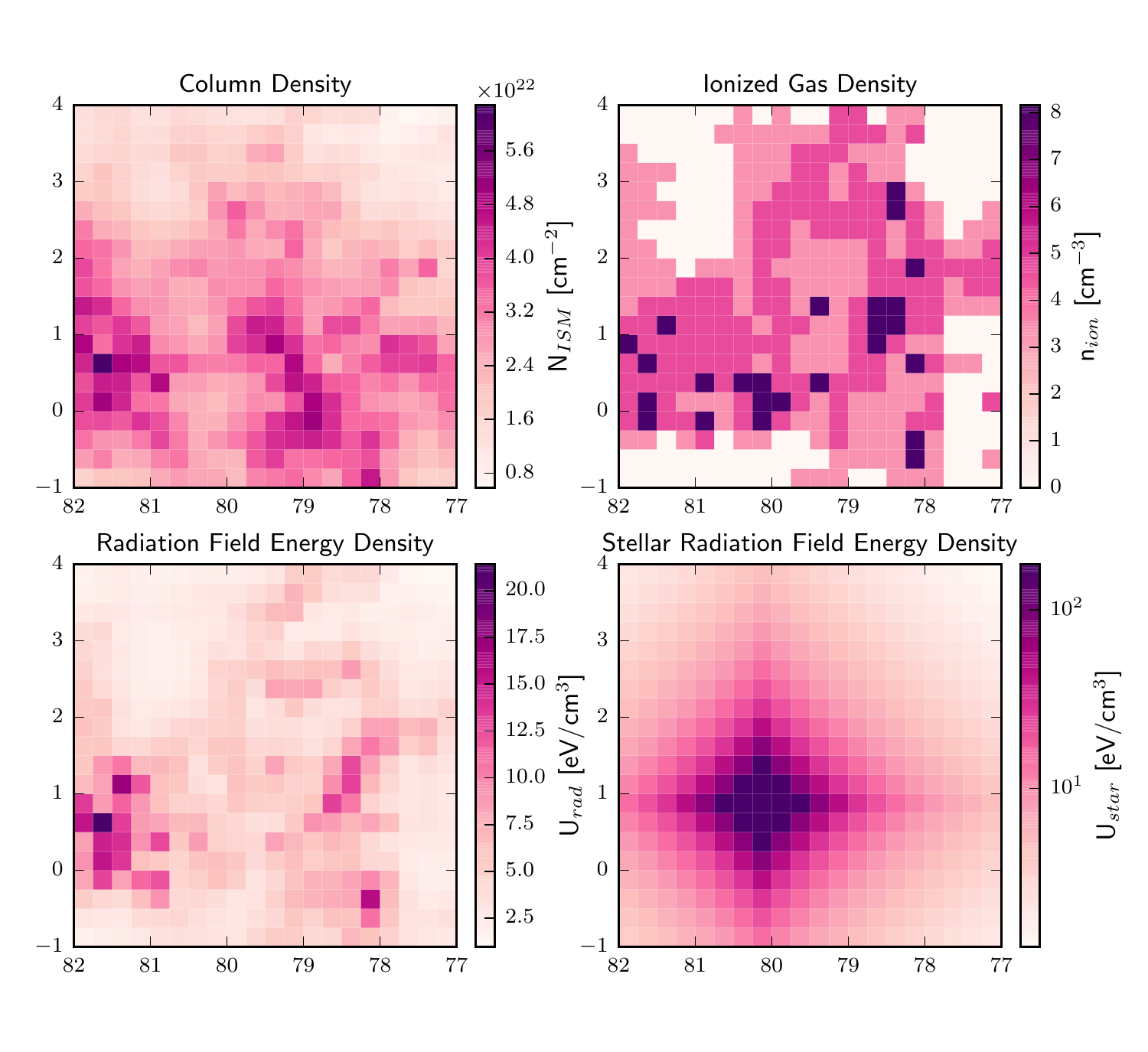}}
\caption{Spatial maps of gas column density (top left), ionized gas volume density (top right), stellar radiation field energy density from Cyg OB2 (bottom right), and infrared radiation field energy density (bottom left).  These maps are used as the main input parameters for the YEGZ models.}
\end{figure*}
%
%

\subsection{Observational Inputs}

The diffuse $\gamma$-ray and neutrino fluxes from pion decay in Cygnus~X primarily depend on the density of interstellar protons, $n_{\text{ISM}}$, and the energy density in cosmic rays, $U_{\text{CR}}$. Our model assumes that the cosmic ray spectrum observed at the Sun [see Eq.~(\ref{proton})] pervades Cygnus~X. Given $U_{\text{CR}}$, the $\gamma$-ray and high-energy neutrino fluxes could be accurately derived from the distribution of gas in three dimensions, information that is not available. We, therefore, approximate the interstellar gas in Cygnus~X by determining the column density of protons, $N_{\text{ISM}}$ in 0.0625~deg$^2$ pixels; see Fig. 1. Our single-zone YEGZ models are then applied to each pixel; see Fig. 2. 

Note that while the pion and $\gamma$-ray source functions, $q_{\pi}$ and $q_{\gamma}$, are sensitive to the physical depth of the gas column because of their dependence on the average gas density, $n_{\text{ISM}} = N_{\text{ISM}} / d$, the $\gamma$-ray flux is independent of the assumed depth.  Instead, the $\gamma$-ray flux (from neutral pion decay and bremsstrahlung) depends on the column density and the assumed angular size of the pixels as 
\begin{align}
dN/dE &= q_{\gamma}(E_{\gamma}) V / 4 \pi D^{2} \propto n_{\text{ISM}} V / D^{2} \nonumber \\
&\propto N_{\text{ISM}} / d \times l^{2} d / D^{2} \nonumber \\
&\propto N_{\text{ISM}} l^{2} / D^{2} \propto N_{\text{ISM}} \theta^{2},
\end{align}
where $D = 1.3$ kpc is the distance to Cygnus~X, $l \approx 5.67$~pc is the length of the side of each 0.25 deg $\times$ 0.25 deg pixel, and $d \approx 113$~pc is the assumed depth of the region which is equal to an angular diameter of 5 deg at the assumed distance.

Measurements of $N_{\text{ISM}}$ are subject to a number of uncertainties. For this calculation, we derive $N_{\text{ISM}}$ in the neutral and molecular interstellar media from optical depth maps obtained by the \textit{Planck} collaboration. We use this information in the form of estimates of the interstellar extinction color excess $E(B-V)$ that depends on the optical depth and, thus, column density of interstellar dust. Following \citep{Draine11}, we assume a uniform ratio of dust-to-gas and adopt $N_{\text{ISM}} / E(B-V) = 5.8 \times 10^{21}$~cm$^{-2}$. We used this ratio to convert the \textit{Planck} $E(B-V)$ maps to a mean $N_{\text{ISM}}$ map for each $0.25 \times 0.25$ deg$^2$ pixel in the Cygnus~X region; see Fig. 1. Our value for $N_{\text{ISM}} / E(B-V)$ is a compromise that is 1$\sigma$ higher than the \textit{Planck} Galactic mean value. The \textit{Planck} mean value, however, does not include ``dark'' molecular gas, i.e. molecular gas not detected via microwave emission from CO, and therefore is a lower limit to the true ratio. 

The distribution of photoionized gas is derived from thermal radio emission maps in \citep{Xu13}. These were placed on our pixel grid via a simple visual estimation from the published figures; see Fig. 1. We converted the observed thermal brightness temperatures to an emission measure following standard techniques. The mean density of ionized gas was found by assuming the H\textsc{II} gas is distributed over a depth of $d \approx 113$~pc in Cygnus~X with a gas filling factor of $\epsilon = 0.5$. This approach is adequate as the ionized gas has only a small affect on the results from our model. 

Cosmic ray interactions with the radiation fields within Cygnus~X also contribute to the production of diffuse $\gamma$-rays. The energy density from thermal emission by interstellar dust grains in the far infrared spectral region, $U_{\text{FIR}}$, is estimated from the 100~$\mu$m intensity maps obtained with IRAS. The intensity observed in each pixel was converted to a mean energy density by adopting a uniform radiation temperature of 25~K in combination with the modified black body radiator model in \citep{Casey12}; see Fig. 1. 

The distribution of direct radiation from the Cygnus OB2 stellar association was calculated assuming a stellar population with stellar ages of $3-4$~Myr and a stellar mass of $\approx 3 \times 10^{4} ~ M_{\odot}$ from \citep{Wright15}. Our luminosity estimate for Cyg OB2, $L_{\ast} = 4.7 \times 10^{7} ~ L_{\odot}$, comes from STARBURST99 models relating the ages and masses of stellar populations to total luminosities. We assume that this source is a blackbody, with $T_{\ast} = 20000$~K, whose flux follows an inverse square law. This leads to an overestimate of $U_{\ast}$ in regions such as Cygnus~X with significant levels of dust absorption; however, our models show that even the unattenuated stellar radiation field is not an important source for $\gamma$-rays from inverse Compton.

\section{Results}

\subsection{Diffuse Interstellar Emission}

\begin{figure}
 \subfigure[~YEGZ Model]{
  \includegraphics[width=0.875\linewidth]{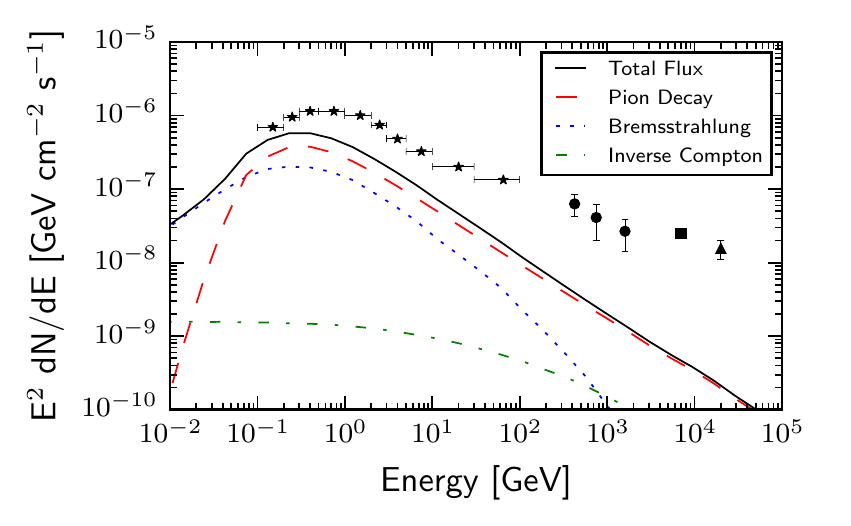}}
 \subfigure[~YEGZ Models with Varying Spectral Indices]{
  \includegraphics[width=0.875\linewidth]{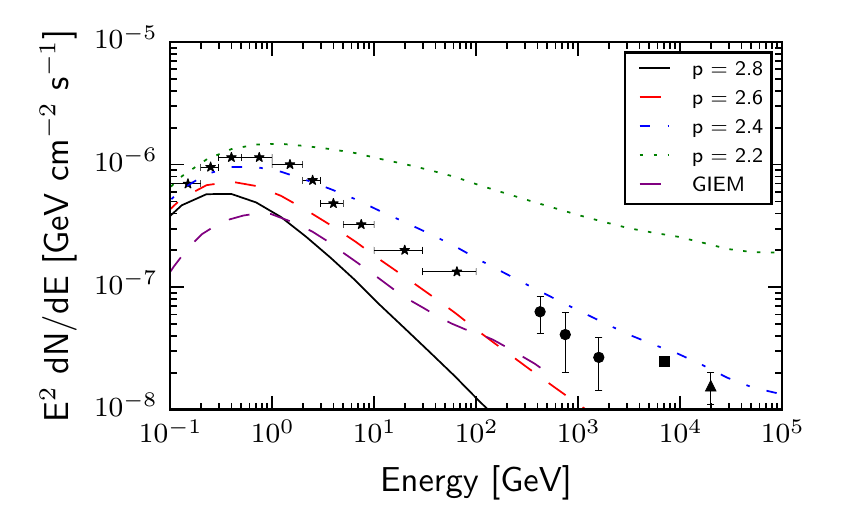}}
 \subfigure[~Residuals]{
  \includegraphics[width=0.875\linewidth]{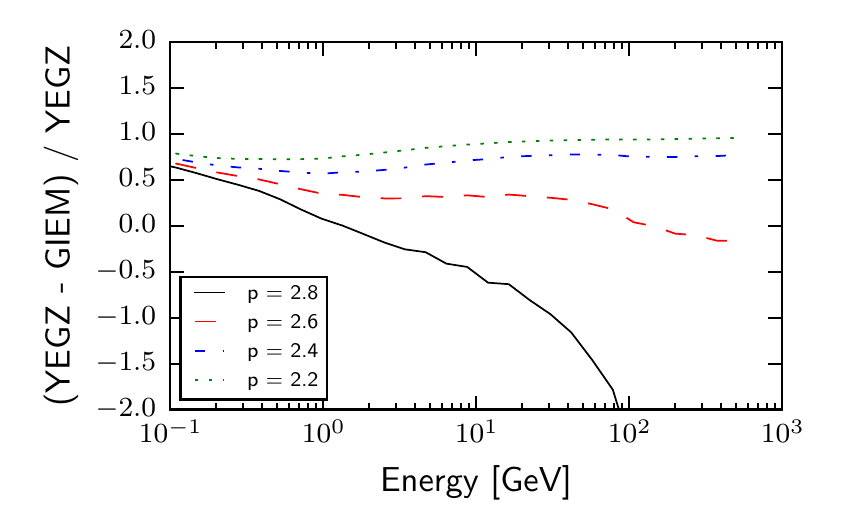}}
\caption{Plots of the total $\gamma$-ray spectrum from the YEGZ models.  Observational data points include data from \textit{Fermi} (black stars \citep{Ackermann12}), ARGO-YBJ (black circles \citep{Bartoli15}), HAWC (black square \citep{Abeysekara17}), and Milagro (black triangle \citep{Abdo07b}).  Top panel: Different components of the $\gamma$-ray spectrum include emission from neutral pion decay, bremsstrahlung, and inverse Compton. Center panel: The total $\gamma$-ray emission is shown for different spectral indices: $p = 2.2 - 2.8$. Bottom panel: Residuals are shown for comparison between the YEGZ models and the GIEM.}
\end{figure}

To model the soft, diffuse interstellar $\gamma$-ray emission in the Cygnus region, we apply the single-zone YEGZ models, using the theoretical framework outlined above and in \citep{YoastHull13,YoastHull14a,YoastHull15}, to each pixel in the maps in Fig. 1.  Summing the fluxes from each pixel, we find that at the lowest energies  ($E \sim 10^{-2}$ GeV) the $\gamma$-ray spectrum is dominated by the flux from bremsstrahlung.  Contributions from neutral pion decay become competitive above $\sim 0.05$ GeV and dominate entirely by $\sim 10$--$100$ GeV, see Fig. 2(a).   $\gamma$-ray emission from inverse Compton is negligible due to the steepness of the cosmic ray electron spectrum at higher energies.

\begin{figure*}
 \subfigure[~YEGZ Model]{
  \includegraphics[width=0.5\linewidth]{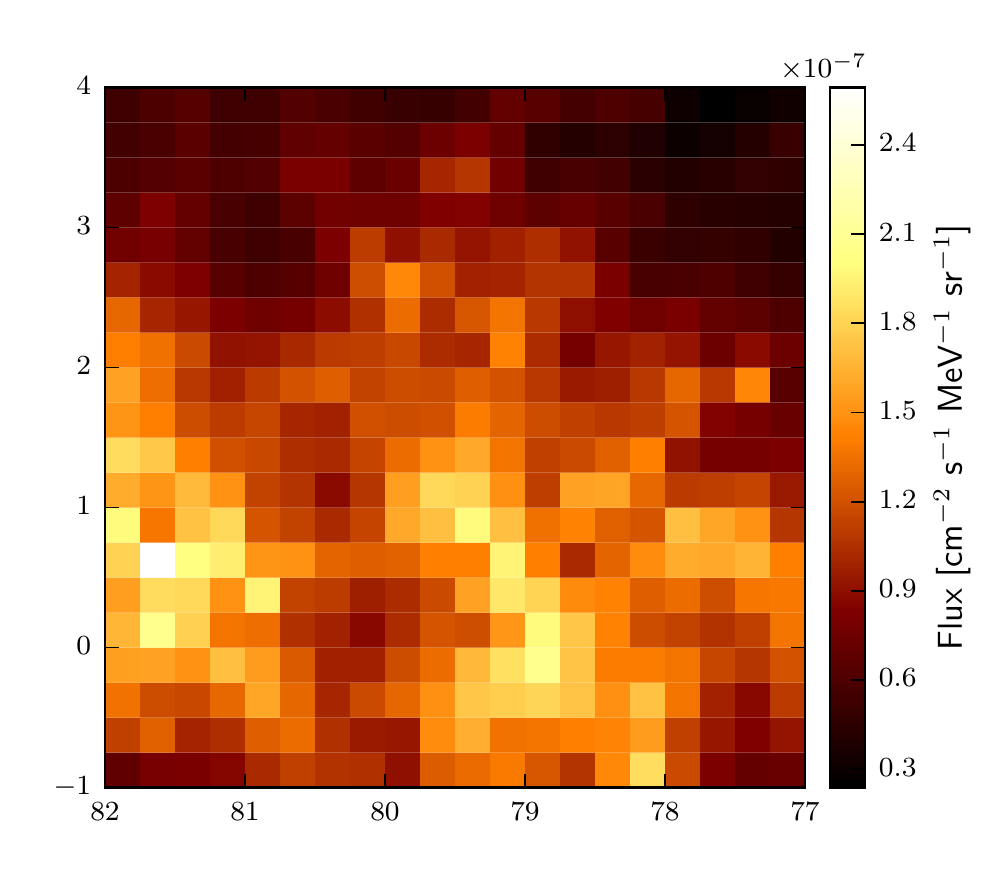}}
 \subfigure[~GIEM]{
  \includegraphics[width=0.4\linewidth]{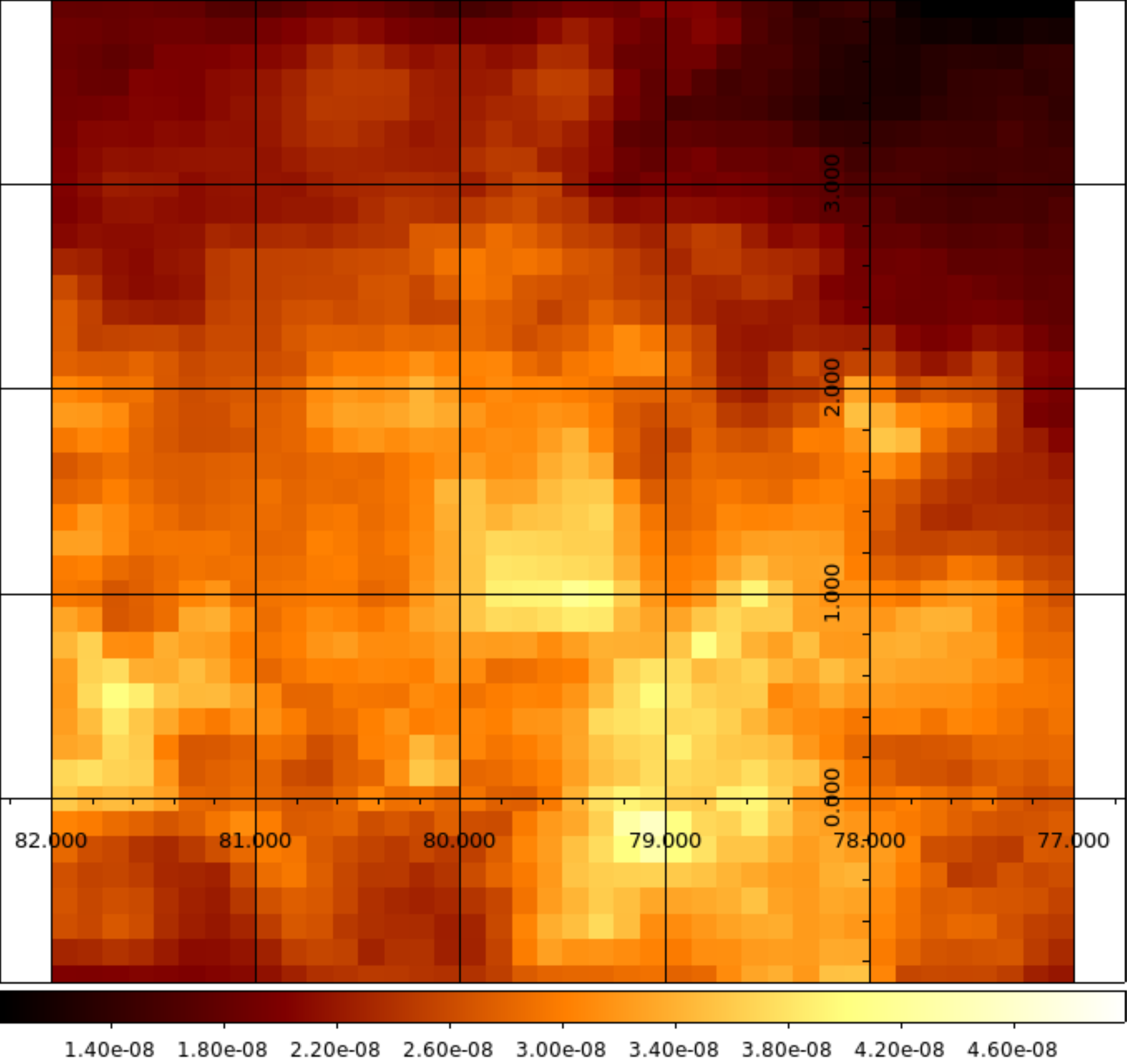}}
\caption{Spatial maps of the modeled $\gamma$-ray flux at 1 GeV.}
\end{figure*}

In addition to assuming a steep ($p = 2.8$) local cosmic ray spectrum, we test cosmic ray spectra with harder spectral indices; see Fig. 2(b).  By varying the second exponent in Eq. (1) from -3.93 to -3.73, -3.53, and -3.33, we test cosmic ray spectra with indices of $p = 2.6, 2.4, 2.2$, respectively.  None of the spectral shapes match well to the observed $\gamma$-ray spectrum, indicating additional sources of emission as expected, and for the harder spectra ($p = 2.2, 2.4$), we overestimate the observed $\gamma$-ray flux entirely.

To compare our YEGZ models to the GIEM adopted by the LAT Collaboration, we downloaded the most recent version available for use with the LAT Pass 8 data, \textit{gll\_iem\_v06.fits}.  The GIEM templates for interstellar gas are derived from spectral line surveys of HI and CO, with corrections for neutral gas and optical depth effects from infrared observations \citep{Ackermann12b}.  These templates are used to calculate $\gamma$-ray emission from neutral pion decay and bremsstrahlung.  In combination with a calculation of the inverse Compton emission from GALPROP simulations, the combined templates are fit to the LAT data in each of 14 logarithmically spaced energy bins from 50 MeV to 50 GeV; for further details, see \citep{Acero16}.

Considering the energy spectra of the different models, we find that the GIEM presents a harder spectrum than our baseline ($p = 2.8$) YEGZ model and has a spectral index closer to $p = 2.6$.  This can be seen in Fig. 2(c), where we plot the differences between the YEGZ models and the GIEM.  The residuals for the GIEM versus of model with $p = 2.6$ is relatively flat and lies the closest to the zero mark.  In contrast, the residuals for $p = 2.2, 2.4$ increase with energy and the residuals for $p = 2.8$ steeply decline with energy.

In addition to looking at the energy spectra, we also extracted maps of the total $\gamma$-ray emission at $\sim 1$ GeV from both our baseline ($p = 2.8$) YEGZ model and the GIEM for comparison; see Fig. 3.  After converting to the same units, we find reasonable agreement of the structures in the $\gamma$-ray emission between the maps.  Further, differences in resolution between the two models account for the larger dynamical range in the total flux in our YEGZ model.

Having established that our YEGZ models for soft, diffuse emission in Cygnus~X are in rough agreement with the GIEM, and, thus, \textit{Fermi} observations, we then calculated the associated neutrino spectrum for the models with spectral indices of $p = 2.6, 2.8$.   The resulting flux is well below the IceCube sensitivity limits for extended sources at 1 PeV \citep{Aartsen14, Aartsen15b}.

\subsection{The Cygnus Cocoon}

In \citep{Ackermann12}, only 11 individual sources were identified in the Cygnus region.  Looking at the third \textit{Fermi} LAT source catalog (3FGL), there are now 24 sources identified within a $4^{\circ}$ radius of the center of Cygnus~X.  These sources include 4 pulsars (PSR), 2 active galactic nuclei (AGN),and  1 SNR with an additional point sources with a potential association with a SNR or PWN, and 16 unassociated (UnID) point sources; see Fig. 4.  To be able to compare our YEGZ models with \textit{Fermi} $\gamma$-ray data for the Cygnus region, we must include these sources from the 3FGL, along with the Cygnus Cocoon and the isotropic $\gamma$-ray background.

We compare the combined $\gamma$-ray spectrum for Cygnus~X with observations from \textit{Fermi}, ARGO-YBJ, HAWC, and Milagro; see Figs. 4 and 5(a).  The spectral fits provided in the 3FGL are only valid between 100 MeV and 300 GeV.  However, for the Cocoon, we extrapolated with spectral fit to higher energies to compare with TeV energy $\gamma$-ray observations.  For the pulsars, we include only the off-pulse emission for the 3 brightest pulsars (J2021+3651, J2021+4026, J2032+4127).  The isotropic $\gamma$-ray background is taken from the LAT background model P8R2\_SOURCE\_V6.

Each of the $\gamma$-ray data points included in Fig. 5(a) were initially given in units of GeV cm$^{-2}$ s$^{-1}$ sr$^{-1}$.  We scaled the \textit{Fermi} \citep{Ackermann12} and ARGO-YBJ \citep{Bartoli15} data down to a region of $5^{\circ} \times 5^{\circ}$ as these observations covered a larger region and the emission outside our selected region is largely negligible.  For the Milagro observations \citep{Abdo07b}, the data originally covered a region with a radius of $1.5^{\circ}$; see Fig. 4, and we scaled the data to a region covering $3^{\circ}$ in radius, equivalent to a box of $5^{\circ} \times 5^{\circ}$. For the HAWC observations \citep{Abeysekara17}, we scaled the data point to a region covering $2^{\circ}$ in radius as their data originally covered a region of only $0.7^{\circ}$ in radius.

\begin{figure}
 \subfigure{
  \includegraphics[width=0.90\linewidth]{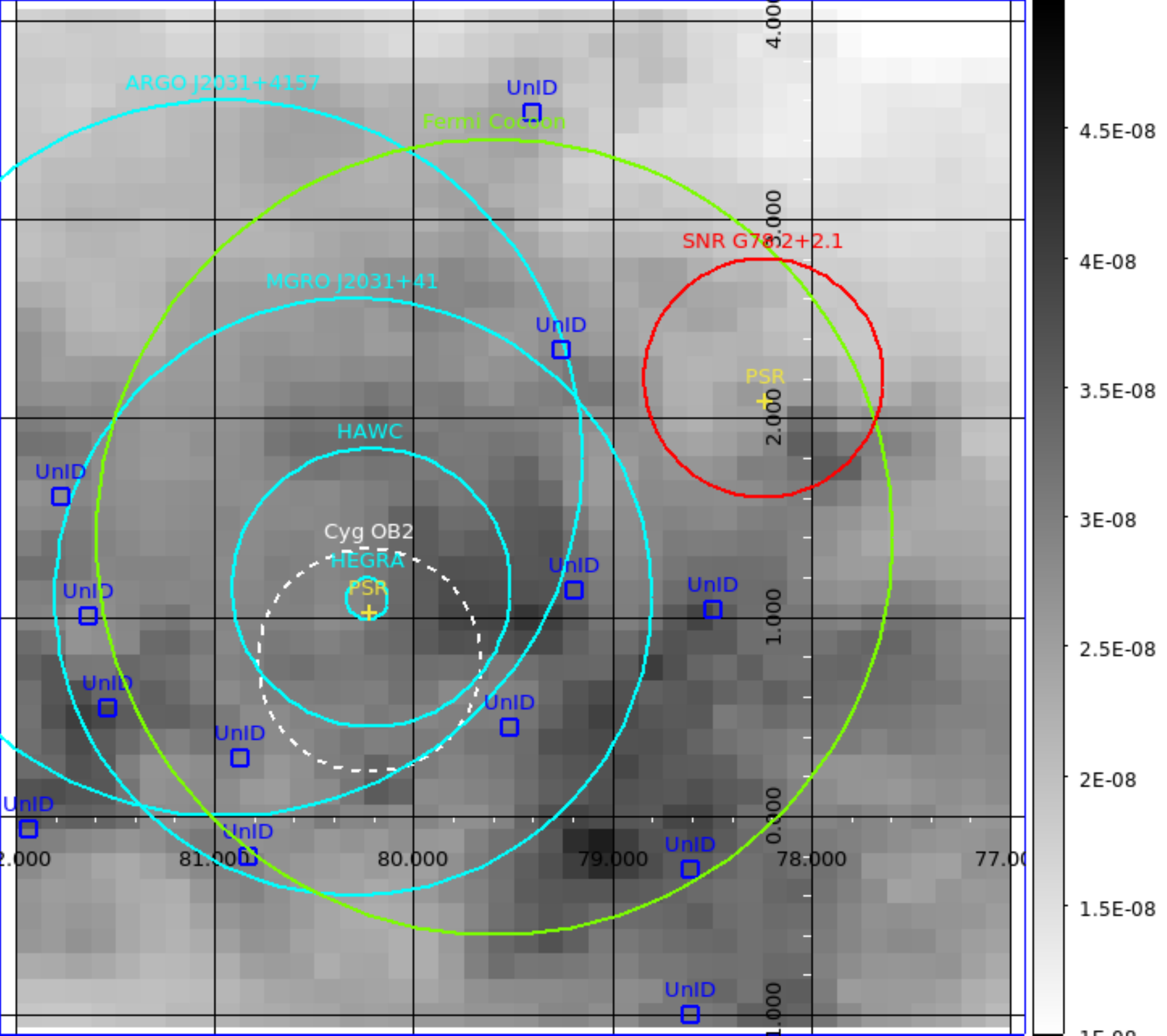}}
\caption{Source map of Cygnus X: GIEM at 1 GeV overlaid with locations of point sources from the 3FGL (UnID - blue square, PSR - yellow cross, SNR - red circle), extended GeV emission (green circle), extended TeV emission (cyan circle), and Cygnus OB2 (white, dashed circle). The regions with extended emission have been fitted with Gaussian sources with free locations and widths (by their respective collaborations).  The best fitting coordinates and areas are shown for ARGO J2031+4157 \citep{Aliu14}, 2HWC J2031+415 \citep{Abeysekara17}, TeV J2032+4130 (HEGRA) \citep{Aharonian02}, Cygnus Cocoon \citep{Ackermann11}, and MGRO J2031+41 \citep{Abdo07b}.}
\end{figure}

Combining the $\gamma$-ray spectrum for the Cocoon, extrapolated to TeV energies, with our modeled diffuse emission and the $\gamma$-ray spectra for point sources in the region gives a total $\gamma$-ray spectrum that is in agreement with both the GeV and TeV energy $\gamma$-ray data; see Fig. 5(a). While we find only rough agreement between our $p = 2.8$ model and the \textit{Fermi} data, we find agreement  between our $p = 2.6$ model and nearly all available data.  This agreement between the models and observations will allow us to use the existing Cocoon spectrum to model further hard neutrino emission from the Cygnus region.

\begin{figure}
 \subfigure{
  \includegraphics[width=0.90\linewidth]{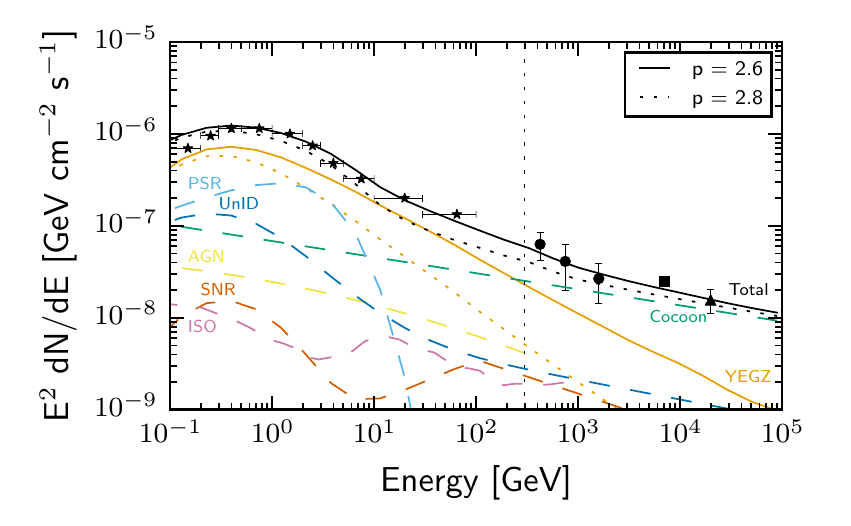}}
 \subfigure{
  \includegraphics[width=0.90\linewidth]{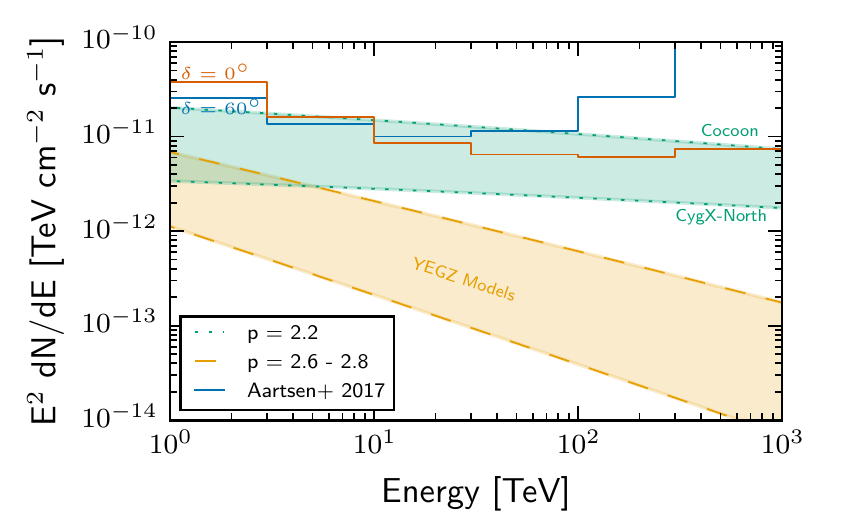}}
\caption{Top panel: plot of the $\gamma$-ray spectra including the YEGZ diffuse model, 3FGL resolved sources, and the Cocoon.  Different components include: YEGZ models, pulsars (PSRs), active galactic nuclei (AGNs), supernova remnants and associated emission (SNRs), unassociated sources (UnID), the isotropic $\gamma$-ray background (ISO), and the Cocoon.  Sources from the 3FGL are valid out to 300 GeV (vertical dotted black line) and extrapolated beyond that. Bottom panel: plot of the neutrino spectra from the soft, diffuse YEGZ models ($p = 2.6 - 2.8$), the Cygnus Cocoon, and the CygX-North molecular cloud complex, along with the point source differential discovery potential for IceCube based on 7 years of data \citep{Aartsen17}. The IceCube sensitivity to extended sources naturally is lower than that for point sources, and thus this plot represents the most optimistic case for detection.}
\end{figure}

To calculate an upper limit on the potential neutrino emission from the Cocoon, we assume that the Cocoon is a single source and is dominated by $\gamma$-rays from neutral pion decay.  Using our single-zone YEGZ interaction model \citep{YoastHull13}, we approximate the spectrum of cosmic-ray protons necessary to reproduce the observed $\gamma$-ray spectrum.  Assuming there is no steepening of the cosmic-ray proton spectrum at higher energies, we find that the neutrino flux ($p = 2.2$) at 1 PeV is a just above the differential discovery potential point sources for IceCube, based on 7 years of data \citep{Aartsen17}; see Fig. 5(b).  As the discovery potential for extended sources should be at least a factor of a few lower (see Fig. 8.1 in \citep{Aartsen15b}), the possibility of detecting the Cocoon is even greater, provided the cosmic ray spectrum is hadronic and extends to PeV energies.

It is likely that several different accelerators and interaction processes produce the hard emission that has been designated the Cocoon, and it is unclear whether the total $\gamma$-ray emission from the Cocoon is dominated by hadronic processes.  As such, we also consider a smaller portion of the Cocoon coincident with a large molecular gas cloud complex which is most likely to be dominated by hadronic emission and could potentially be due to a single, hidden accelerator (a SNR or a PWN).  The region we consider is in CygX-North centered on $(l = 81.5^{\circ},~b = 0.5^{\circ})$ which is to the left of Cyg OB2; see Fig. 7 in \citep{Schneider06}.  

Again, using our single-zone YEGZ model \citep{YoastHull13}, we match a cosmic-ray proton spectrum to the $\gamma$-ray spectrum for this subregion given in the supplementary materials of \citep{Ackermann11}; see Fig. S6.  We find that the neutrino flux at 1 PeV is a factor of $\sim 4$ below IceCube's differential discovery potential \citep{Aartsen17}; see Fig. 5(b).  This indicates that CygX-North is unlikely to be detected by IceCube as a point source and the possibility of being detected as an extended source is slim.

\section{Summary \& Discussion}

In applying our semianalytic YEGZ model to the Cygnus X region, we sought to minimize free parameters.  We calculated the spectra for diffuse $\gamma$-ray and neutrino emission by assuming a local cosmic ray spectrum with spectral indices between $p = 2.6 - 2.8$ and by deriving spatial maps for the gas column density and infrared radiation fields from observations by \textit{Planck} and IRAS.  Checking our YEGZ model for the soft, diffuse cosmic ray population against the GIEM adopted by the LAT Collaboration, we find rough agreement of the flux map and the spectral energy distribution; see Figs. 1 and 2.

When combining the $\gamma$-rays resulting from the soft, diffuse cosmic ray population with the $\gamma$-ray spectra for both point sources and the Cocoon (extended to TeV energies), the total spectrum agrees with observations by \textit{Fermi}, ARGO-YBJ, HAWC, and Milagro.  Based on this agreement between the various cosmic ray populations at TeV energies, we use the $\gamma$-ray emission from the Cocoon to derive a corresponding cosmic-ray proton population (assuming only hadronic emission) and extend the population to PeV energies to calculate an upper limit on the neutrino flux.

While neutrino emission from the diffuse, soft cosmic ray population acting alone in Cygnus X results in neutrino fluxes several orders of magnitude below the current IceCube sensitivity limits, the neutrino flux from a hard cosmic ray population equivalent to that required for the Cocoon results in a flux that is potentially detectable by IceCube.  We also calculated the neutrino emission from the subregion CygX-North within the Cocoon maps which directly onto a particularly dense molecular cloud complex and found that the corresponding flux is a factor of $\sim 4$ below the current discovery potential for point sources.

The Cygnus X region is complex, and the origin of the hard $\gamma$-ray emission component associated with the Cocoon remains unclear. Currently Cygnus X has not been reported as a detection by IceCube. However, if the Cocoon is hadronic and extends to PeV energies with a flat spectrum, then eventual detection of high-energy from the Cocoon is possible with IceCube.  A detection of high-energy neutrinos from Cygnus X would provide important clues to the origin of the Cocoon $\gamma$-ray emission and would cleanly establish the presence of a thus far undetected PeV energy hadronic accelerator in this part of the Galaxy.

\begin{acknowledgments}
We thank the referee for their valuable comments and Dr. Cristina Popescu for providing her model stellar radiation field for Cygnus X that provided a check on our assumptions.  This research has made use of the NASA/IPAC Infrared Science Archive, which is operated by the Jet Propulsion Laboratory, California Institute of Technology, under contract with the National Aeronautics and Space Administration.  This research was supported in part by the U.S. National Science Foundation under Grants No.  AST-0907837, No. ANT-0937462, and No. PHY-1306958 and by the University of Wisconsin Research Committee with funds granted by the Wisconsin Alumni Research Foundation.
\end{acknowledgments}

\normalem

%
\end{document}